# ScAtt: an Attention based architecture to analyze Alzheimer's disease at cell type level from single-cell RNA-sequencing data

Xiaoxia Liu[1#], Robert R Butler III[1], Prashnna K Gyawali[1], Frank M Longo[1], Zihuai He[1,2#]

[1]Department of Neurology and Neurological Sciences, Stanford University, Stanford, CA 94305, USA
[2]Quantitative Sciences Unit, Department of Medicine, Stanford University, Stanford, CA, 94305, USA

# Correspondence to: Xiaoxia Liu (xxliu@stanford.edu), Zihuai He (zihuai@stanford.edu)

**Abstract**

Alzheimer's disease (AD) is a pervasive neurodegenerative disorder that leads to memory and behavior impairment severe enough to interfere with daily life activities. Understanding this disease pathogenesis can drive the development of new targets and strategies to prevent and treat AD. Recent advances in high-throughput single-cell RNA sequencing technology (scRNA-seq) have enabled the generation of massive amounts of transcriptomic data at the single-cell level provided remarkable insights into understanding the molecular pathogenesis of Alzheimer's disease. In this study, we introduce ScAtt, an innovative Attention-based architecture, devised specifically for the concurrent identification of cell-type specific AD-related genes and their associated gene regulatory network. ScAtt incorporates a flexible model capable of capturing nonlinear effects, leading to the detection of AD-associated genes that might be overlooked by traditional differentially expressed gene (DEG) analyses. Moreover, ScAtt effectively infers a gene regulatory network depicting the combined influences of genes on the targeted disease, as opposed to examining correlations among genes in conventional gene co-expression networks. In an application to 95,186 single-nucleus transcriptomes from 17 hippocampus samples, ScAtt shows substantially better performance in modeling quantitative changes in expression levels between AD and healthy controls. Consequently, ScAtt performs better than existing methods in the identification of AD-related genes, with more unique discoveries and less overlap between cell types. Functional enrichments of the corresponding gene modules detected from gene regulatory network show significant enrichment of biologically meaningful AD-related pathways across different cell types.

# Introduction

Alzheimer's disease (AD) is a severe neurodegenerative disease that impacts a variety of cell types and disrupts multiple interrelated systems in the brain, resulting in large scale neuronal death and cognitive decline into dementia[1,2]. AD is currently the sixth leading cause of death and affects over 35 million people globally[3]. Brain pathologies associated with AD can start 10–20 years prior to the onset of dementia symptoms[4,5]. Even though there are currently no disease-modifying treatments for AD at the moment, early diagnosis could lead to early therapeutic interventions that slow the disease progression over time, help tailor disease management and plan future care, and improve the quality of AD patients' life[6,7]. Understanding the pathogenesis of this disease can not only aid in early diagnosis but also drive the development of new targets and strategies for the prevention and treatment of AD[8].

Single-cell RNA-sequencing (scRNA-Seq) techniques are extremely useful for investigating tissue heterogeneity, revealing differentiation dynamics, assessing transcriptional stochasticity, and dissecting the function/dysfunction of highly heterogeneous cells at the single-cell level. Thus, the corresponding scRNA-seq data analyses can significantly improve our understanding of the disease. In recent years, various computational tools for the analysis of scRNA-seq data have been developed[9-13]. These tools are intended to capture gene features and extract information from scRNA-seq data in order to predict cell types[13], find differentially expressed genes (DEGs) [14,15], infer gene relationships[16], construct gene networks[17] and predict disease states[18]. However, existing DEG finding methods such as Seurat[15] perform a marginal association test in a linear model, which is unable to capture the nonlinear relations among genes. Also, prior studies have indicated that AD does not affect all cell types equally[19], implying that some are more vulnerable than others. In order to investigate cell-type-specific AD-related changes, it is necessary to examine scRNA-seq data at the cell-type level between patients and healthy individuals. Further, inferring gene regulatory networks from cell type expression could identify altered mechanisms of AD at the functional pathway level. Through network analysis, it is possible to investigate how genes share biological functions and how disease susceptibility propagates their information through interactions. However, existing approaches for gene regulatory network construction from scRNA-seq data only explore gene-gene correlation, they do not examine the correlation of expression changes directly related to the disease of interest.

In this paper, we propose an end-to-end Attention based architecture for analyzing scRNA-seq data (ScAtt). We used Attention-based Transformer architecture to aggregate global and local level information from the data and learn linear and nonlinear features from the data. Attention-based Transformers have achieved superior performance in many domains, including vision and natural language processing[20-23]. The value of a Transformer model hinges not only on its accuracy but also on its high interpretability enhanced by the Attention mechanism. We train our models on scRNA-seq data with cell state classification for two conditions (AD and Control) as the objective to validate model fitting. Learning RNA expression signatures for individual cells enables ScAtt to effectively integrate gene expression into its parameter matrices. ScAtt can efficiently classify the cell state, showing that the model could effectively learn from gene expression data. Thus, by interpreting the parameters of the trained models, our architecture can then extract AD-related genes and construct a regulatory network for AD-related genes based on feature associations. Additionally, to investigate cell-type-specific AD-related changes, we separately train ScAtt on scRNA-seq data from fourteen different cell types and obtain cell-type-specific AD-related genes and regulatory networks.

In summary, we introduce ScAtt, a novel Attention-based architecture specifically designed for analyzing scRNA-seq data at the cell type level. The capabilities of ScAtt include the identification of cell-type-specific AD-related genes by utilizing the model parameters learned from data from both disease and control states of each cell type. Moreover, it enables the construction of cell-type-specific regulatory networks with AD-related genes as nodes by using feature representations derived from model that account for disease states. In addition, we also provide biologically meaningful modules of correlated AD-related genes detected from constructed regulatory networks for further analysis. Overall, we propose a new deep-learning approach to analyze scRNA-seq data that enables studying Alzheimer's disease from multiple perspectives.

**Results**

**Overview of ScAtt for analyzing scRNA-seq data.**

In this paper, we proposed an Attention-based architecture, ScAtt, to analyze scRNA-seq data at the cell type level. Our approach utilizes scRNA-seq data as input and the cell states of AD patients or controls as the output, thereby validating the model's fitting and enabling the exploration of disease-related genes and the inference of gene regulatory networks. Figure 1a presents a workflow summary of ScAtt, illustrating its three main components: data processing, model framework, and model training. The process begins with the preparation of scRNA-seq data. Specifically, we utilized data from 95,186 single-nucleus transcriptomes across 17 hippocampus samples, including 8 controls and 9 AD cases, stratified into fourteen cell types. These cell types include Venous endothelial cell (Venous), T cell, Smooth muscle cell (SMC), Pericyte, Capillary endothelial cell (Capillary), Arterial endothelial cell (Arterial), Oligodendrocyte (Oligo), Perivascular fibroblast (P. Fibro), Ependymal cell (Ependymal), Microglia, Astrocyte, Oligodendrocyte progenitor cell (OPC), Meningeal fibroblast (M. Fibro), and Neuron. For our datasets, six AD and five healthy controls contributed 61,113 cells (30,979 AD; 30,134 control) for training, one AD and one healthy control provided 12,088 cells (5,320 AD; 6,768 control) for validation, and the remaining two AD and two healthy controls yielded 21,985 cells (10,387 AD; 11,598 control) for testing. ScAtt models were separately trained on scRNA-seq data from each cell type to explore AD-related genes at the cell type level. We employed the parameters of the trained model for each cell type to measure feature importance scores, which were then used to identify cell-type-specific AD-related genes (as depicted in Figure 1b). Additionally, we utilize the feature embeddings obtained from ScAtt to calculate gene relations, which enable us to construct cell-type-specific gene regulatory networks related to AD as illustrated in Figure 1c. Subsequent analyses, including module detection and enrichment analysis, were performed on these constructed regulatory networks, providing further insights into the molecular mechanisms of AD.

**ScAtt learned multiplex cell-type specific patterns missed by conventional methods.**

Machine learning algorithms aim to learn a mapping function that maps input variables to output prediction. As the model is trained for predicting cell state using scRNA-seq data, if a model can accurately predict whether a cell is from AD or healthy people, it indicates that the model observed the differences between AD patients and the control cohorts. In the following, we evaluate ScAtt in terms of model performance for predicting cell state, feature learning capacity, model stability, and scalability.

*Strategies for training models.* We benchmarked ScAtt against established machine learning algorithms, including logistic regression[24], support vector machine[25] (SVM), decision tree[26], Adaboost[27], and XGBoost[28], which are widely used in various biomedical applications. Additionally, we included Residual Network[29] (ResNet) and Deep Neural Network[30] (DNN) methods as representatives of deep learning approaches in our comparison. These deep learning methods are notably recognized for their ability to identify complex patterns and extract high-level features. Figure 2a shows the different training strategies between baseline models and our method. The baseline methods typically analyze gene expression data on a per-cell basis, generating an output that indicates the state of each individual cell. These methods primarily focus on the relationships between genes within the same cell under a specific cell type, but do not extend this analysis to gene interactions across cells within the same cell type. In contrast, ScAtt utilizes scRNA-seq data from all samples within each cell type, employing an Attention-based Transformer framework to explore gene-gene expression associations. This approach allows ScAtt to not only consider the gene-gene relationships within individual cells of the same type but also to examine gene-gene interactions across different cells within that cell type. By training on all available scRNA-seq data for each cell type, ScAtt can exploit local features (gene-gene relations per cell) and global features (gene-gene relations across cells) of the same cell type to speculate cell states. The detailed training process of all baseline methods and ScAtt is described in the Methods.

*ScAtt learns nonlinear patterns from scRNA-seq data.* Figure 2b shows the AUC performance of all competing methods. It shows that most baseline machine learning methods can accurately classify cell states using scRNA-seq data. Logistic regression, SVM (with linear kernel), AdaBoost, XGBoost, ResNet, DNN and ScAtt all achieved AUCs greater than 0.80 on Astrocyte, Pericyte, Capillary, Venous, SMC, Arterial, Microglia, and P. Fibro cell types, demonstrating the potential of various machine learning methods to differentiate AD cell state across multiple cell-types. Detailed results of other evaluation criteria can be found in Supplementary Table 1. Our model outperforms all competing models on the majority of cell types, with an identification ability of up to 0.976 AUC on Pericyte cell type. Notably, the enhanced performance of ScAtt over linear models like logistic regression and SVM suggests its proficiency in capturing nonlinear effects present in scRNA-seq data. On the other hand, decision tree models generally underperform; for instance, their AUCs on Oligo, Astrocyte, Venous, OPC, Microglia, P. Fibro, Ependymal, and Neuron are all below 0.60. This significant difference from ScAtt's consistently high AUC scores, often above 0.80, may be due to decision trees' inability to model the quantitative and continuous nature of scRNA-seq data. From Figure 2b, it is observed that ScAtt's performance on M. Fibro and T cells is not optimal, likely due to the limited amount of training data available for these cell types. The data distribution of all the training data sets (the lower right corner of Figure 2b) shows that less than 1% of cells in the entire scRNA-seq training set are from M. Fibro cells (124 cells) and T cells (114 cells) cell types. This scarcity of data can impact the efficacy of deep learning models like ScAtt, as evidenced by the similarly poor performance of DNN and ResNet in these small datasets. Nonetheless, ScAtt still achieves reasonable performance in these two cell types, with AUCs of 0.774 and 0.624, respectively, and only AdaBoost slightly outperforming ScAtt on T cells. In addition, after obtaining the classification results at the cell level, we could use the results to gain the disease state at the patient level, which is shown in the supplementary material. This extends the utility of our model from cellular analysis to potentially offering insights into patient-specific disease profiles.

*ScAtt learns Cell-type-specific patterns.* Given that different brain cell type impact different mechanisms in AD processes, our independent cell-type models intend to obtain cell-type-specific differential genes. To validate the cell-type-specific patterns learned by our model, we conducted a cross-cell-type evaluation, as depicted in Figure 2c. In this evaluation, models trained for a specific cell type were used to classify cell states of other types. If a model accurately learns characteristics unique to the cell type it is trained on, it should exhibit superior performance on that cell type compared to others. Conversely, similar performance across different cell types would indicate that the model has extracted more general, invariant patterns applicable to various cell types. The results of all pairwise cross-cell-type evaluations are displayed in Figure 2c, where the value denotes the AUC performance of a model trained on the row cell type and tested on the column cell type, with the best performances per row highlighted in red boxes.

The experimental results demonstrate that different types of cells have different gene expression patterns. Figure 2c shows that all the competing approaches are capable of learning the distinct gene expression patterns for various cell types. For example, logistic regression, SVM, AdaBoost, XGBoost, ResNet, DNN learned cell-type-specific features for various cell types indicated by the number of red boxes on the diagonal. Particularly, logistic regression and XGBoost learned cell-type-specific patterns for more than half of all cell types. ScAtt outperforms all other methods in cell-type-specific pattern learning, as shown by the highest number of red boxes on the diagonal in Figure 2c. The results show that ScAtt learned unique patterns for Oligo, Astrocyte, Pericyte, Capillary, Venous, SMC, Arterial, Microglia, P. Fibro, and M. Fibro. Figure 2b and 2c collectively suggest that the baseline methods' capacity to predict cell states correlates with the ability to learn specific cell-type patterns. The better a model learns gene expression patterns for a particular cell type, the better its performance predicts cell states for that cell type. For example, ScAtt learned the specific features for most cell types (10 out of 14 cell types) and correspondingly achieved the best performance on cell state prediction. Similarly, the logistic regression method achieved the second-best performance in specific-cell-type feature learning. Accordingly, it was the second-best method in terms of its performance in cell state prediction. In contrast, Figure 2c shows that the decision tree approach performs poorly in learning cell-type-specific information (2 out of 14 cell types) and thus shows the lowest performance in cell state prediction. Additionally, our comparisons with ResNet and DNN reveal that leveraging the Transformer architecture to learn both global and local features not only enhances model performance but also captures cell-type-specific patterns more effectively. In conclusion, the results in Figures 2b and 2c demonstrate that ScAtt not only achieves the highest AUC performance but also learns more cell-type-specific patterns than any other method tested.

*The stability and scalability of scAtt.* Hyperparameters in machine learning methods are crucial as they dictate the learning process and significantly impact model performance. In the realm of deep learning, it is generally observed that larger models outperform their smaller counterparts. ScAtt, our model, is primarily regulated by three hyperparameters: the hidden dimension, the number of Transformer encoders, and the number of attention heads. Since the model size does not relate to the number of attention heads (refer to the Methods section for more details), the stability analysis here focuses on the impact of the other two hyperparameters (hidden dimension and number of encoders) on model performance and size. Figure 2d shows the performance and size of the models trained on two cell types: Venous and P. Fibro, as examples (with additional data in Supplementary Figure 6). From the figure 2d, we can notice that the model size correlates directly with the hyperparameters, increasing linearly as the hidden dimension and number of Transformer encoders are augmented. For instance, models for Venous and P. Fibro with identical hyperparameter configurations exhibit the same size. The largest model, featuring three Transformer encoders and 512 hidden dimensions, comprises 82.05M parameters, while the smallest, with one encoder and eight hidden dimensions, has just 0.85M. Even though performance tends to improve with increasing model size. However, it is noteworthy that the smallest models still deliver reasonable performance, despite being 100 times smaller than the largest models. For example, the smallest models for Venous and P. Fibro (0.85M) both achieve AUC scores above 0.75. This consistent performance across a range of hyperparameters allows for flexible model selection based on available data and hardware resources. In this paper, we use the largest model size as the experimental model for subsequent analysis.

**Enhanced Detection of AD-Related Genes by ScAtt Compared to Traditional Models**

The previous results show that ScAtt has superior AUC and could learn cell-type-specific patterns. Here, we further analyze the ability of ScAtt in detecting AD-related genes. The feature importance score assigned to each feature reflects the relative importance of each gene feature in our model for classifying the cell state. Genes with a higher feature importance score demonstrate a stronger ability to distinguish between AD and Control conditions. Consequently, these genes may represent the differential genes between the two states of AD and Control. These cell-type-specific differential genes will enable us to identify critical phenotype-determining variables in health and disease systems. A weighted value for each feature could be calculated using the internal weight parameters of the trained model for each cell type. The weighted value reflects the contribution of the gene feature in our model (see Methods for the detailed calculation of weighted values). Accordingly, the genes can be ranked in descending order by their weighted values in terms of their association with AD. The higher the weighted value is, the more important the gene is in determining the cell state and the more likely it is an AD-relevant gene.

*Comparison with alternative machine learning methods.* For comparison purposes, we used 45 known AD-related GWAS genes[31] and 584 Treat-AD nominated genes[32] as target genes (complete gene lists are available in the supplementary file). We evaluated the capability of ScAtt and other machine learning methods in detecting these known genes. We utilized the python sklearn package[33] to calculate the feature importance for machine learning methods and we followed the same methodology employed in ScAtt to determine the feature importance for ResNet and DNN. In our comparison, the top K genes, as per their ranking, were selected, and all the known genes identified for each cell type were aggregated to calculate the total number of discovered genes per method. Figure 3a presents the comparison results with different machine learning methods. The x-axis indicates the threshold of K, which determines the number of genes being selected, while the y-axis shows the total number of known AD-related genes identified by each competing methods among the selected genes. According to Figure 3a, ScAtt consistently identified more known genes compared to the other seven methods, regardless of whether the target gene sets were GWAS or Treat-AD. Notably, decision trees and AdaBoost were the least effective, performing worse than other methods across any K threshold. The experimental results validate the learning ability of ScAtt using only scRNA-seq data to recapitulate previously identified AD-risk genes[31,32] that have been discovered through combined multi-omics approaches.

*Comparison with marginal association test in linear models.* Seurat[15] is a widely-used tool for differential expression gene (DEG) analysis in scRNA-seq data. We conducted a comparative analysis of the gene importance rankings from ScAtt with the DEG results obtained by Seurat for each cell type. Specifically, we assessed the number of AD-related genes identified through the gene feature importance derived from ScAtt compared to the DEG results obtained using Seurat. For each cell type, the DEG analysis was performed using MAST algorithm[34] with Seurat, comparing AD and control cohorts. We then matched the number of genes identified by Seurat at a specific p-value cutoff with those ranked by ScAtt based on feature importance for each cell type. Finally, all the genes found for each cell type were combined to obtain the final total discovered known target genes. Figure 3b reveals that ScAtt consistently identifies more AD-related genes than Seurat across various p-value thresholds, irrespective of whether the genes were part of the AD GWAS or Treat-AD gene sets. Furthermore, Figure 3c illustrates the overall discovery of GWAS and Treat-AD genes by both ScAtt and Seurat at a p-value threshold of 0.05. As shown in Figure 3c, ScAtt based on feature importance score found 43 out of 45 GWAS genes, while Seurat found 35 genes; ScAtt found 542 genes out of 584 Treat-AD genes, while Seurat found 493 genes. This suggests that ScAtt can discern patterns overlooked by marginal linear models like Seurat.

In addition to the total gene count, we also examined the cell-type specificity of ScAtt ranked genes by comparing gene overlap between different cell types. Figure 3d depicts the intersection heatmap of the AD-related genes identified with ScAtt and Seurat across various cell types. The pairwise intersections of genes found in cell types by ScAtt are substantially lower than the intersections of genes found by Seurat, suggesting that our non-linear model is capturing not only more AD-relevant genes, but more cell-type specific genes that define unique cellular AD pathologies. Given the tendency for immune and glial cell types to dominate whole tissue bulk RNAseq differential results[35], this is encouraging together with high cell state classification AUCs like that of Pericytes.

We also examined the results obtained when using the adjusted p-value of Seurat. The adjusted p-value is obtained based on Bonferroni correction using all features in the dataset. Following the same procedure, the adjusted p-value was used to determine the number of genes selected for each cell type, and then we analyzed how many known GWAS or Treat-AD nominated genes were discovered by each method. The discoveries of known GWAS or Treat-AD genes by both ScAtt and Seurat using various adjusted p-value thresholds are detailed in Supplementary Figure 1a. Supplementary Figure 1b shows the total number of GWAS and Treat-AD genes found by ScAtt and Seurat with an adjusted p-value cutoff of 0.05. As expected, fewer genes were identified using the adjusted p-value compared to the unadjusted p-value, as evident from comparing Figure 3c with Supplementary Figure 1a. Nonetheless, ScAtt consistently identified more AD-related genes than Seurat across different adjusted p-value thresholds, regardless of the gene sets used. The cell-type specificity analysis, using the adjusted p-value cutoff, aligns with the findings in Figure 3d, where ScAtt discovered more unique AD-related genes specific to each cell type.

*The top-ranked genes based on feature importance*. The top five genes based on feature importance score for each cell type are listed in Table 1. We conducted a focused analysis on several top-ranked genes that are not featured in the AD-related GWAS or Treat-AD gene datasets. For the Oligo cell type, the ADAMTS18 gene is the top-ranked gene, and a previous study indicates that ADAMTS18 is upregulated in the oligodendrocytes of Alzheimer's disease patients[36]. In SMC and Capillary cell types, ADAMTS9 is identified as the highest-ranked gene. It also ranks as the second and third highest in T cell and Arterial cell types, respectively. Multiple studies have established a connection between ADAMTS9 and Alzheimer's disease[37-39]. NEAT1 is the top-ranked gene for T cells, and some studies suggest that the long-non-coding RNA NEAT1 may be a potential target gene for Alzheimer's disease[40-42]. In three cell types –P. Fibro, Pericyte, and Venous – the XIST gene ranks highest. Many studies reveal that XIST plays a role in AD and a more recent study found that lncRNA XIST induced Aβ accumulation and neuroinflammation by the epigenetic repression of NEP in AD[40]. VEGFA is the top ranked gene for OPC cell type and researchers suggest it may be a key target for potential therapies against Alzheimer's disease[43,44]. The discovery of these genes, which are closely linked to Alzheimer's disease, not only highlights their potential significance in the pathology and treatment of AD but also underscores the capability of ScAtt in identifying AD-related genes.

**Constructing regulatory networks for AD-related genes**.

Inferring gene regulatory networks from cell type expression could identify altered mechanisms of AD at the functional pathway level. Through network analysis, it is possible to investigate how genes share biological functions and how disease susceptibility propagates their information through interactions. Many methods for mining gene-gene relationships from expression data have been developed, including techniques based on correlation, mutual information, and correlation, mutual information, and deep learning based methods for co-expression analysis[16,45–48]. These methods form a foundational step in the construction of gene-gene networks. However, they predominantly focus on modeling gene-gene correlations without directly linking to specific cell types or the targeted disease. Our proposed framework, in contrast, provides a latent representation for each feature within a specific cell type. These representations enable us to quantify gene relationships and construct cell-type-specific regulatory networks, where genes are positioned as nodes and their relationships as edges. This approach goes beyond mere calculation of gene co-expression. It incorporates the genes' combined impact on cell state determination. In other words, we leverage the information from different disease conditions to construct the regulatory network, so an edge between two genes means they are more likely to be closely related in function for differentiating the two disease conditions. Detailed methodology for constructing these networks is elaborated in the Methods section.

*Regulatory network constructed via ScAtt is functionally enriched in AD-related pathways*. A significant challenge that arises when evaluating gene-gene network construction algorithms for scRNA-seq data is the lack of a "ground truth", such as AD-related gene and gene interactions[49]. So, we adopted an alternative approach by investigating whether the constructed regulatory network is enriched for a curated functional ontology of Alzheimer's Disease. We used the "Alzheimer's Disease" pathway from KEGG (hsa05010) to compare our network with three benchmark networks: an AD Co-expression Network using gene expression data from AD patients, a Control Co-expression Network using data from control cohorts, and a Contrast Co-expression Network, highlighting differences between the AD and Control networks. It was reported that linear similarity measures like Pearson's correlation coefficient (PCC) are suitable for exacting the correlated pattern from gene expression data[50], and we used the PCC score to construct the three benchmark Co-expression networks (details in the Methods sections). Figure 4a provides a schematic overview of this network construction process. Figure 4b displays the enrichment results of the Alzheimer's Disease pathway in these four networks for each cell type. All the enrichment analysis of KEGG in this paper were carried out using KOBAS[51]. Additionally, we looked at other KEGG pathways and the corresponding results for all the networks are shown in the Supplementary materials. Within each network, we identified the number of genes overlapping with the "Alzheimer's Disease" pathway (# overlapped gene) and the total number of nodes in the networks (# genes in network). We then employed the Enrichment Ratio (ER), calculated as ER = (# overlapped gene / # total gene in the pathway) / ( # genes in network / # total gene), as a metric to compare the enrichment results across networks. As shown on the y-axis of Figure 4b, the ScAtt networks across all fourteen cell types exhibit the highest enrichment when compared to the benchmark networks. Additionally, the Alzheimer's Disease pathway is significantly enriched in the ScAtt networks for all cell types, except M. Fibro and T cell, as indicated by the adjusted p-values shown above the bars in Figure 4b. The ScAtt-constructed regulatory networks show greater functional enrichment in AD pathways than those created through co-expression. This is attributed to ScAtt's ability to capture gene-gene relationships that consider the joint effect of genes on cell state determination, closely relating to AD status. The enrichment analysis confirms that the networks constructed by ScAtt are relevant to Alzheimer's Disease, providing insights that go beyond merely identify AD-related genes via ScAtt based on their feature importance scores. This network analysis offers a more comprehensive understanding of AD.

*The modules detected from the constructed regulatory networks via ScAtt are functionally enriched in AD-related pathways.* The networks were constructed by leveraging disease and control status across all the cells within a specific cell type, at which point gene-gene relationships can be discovered through module detection. The details of the module detection method we used are described in the Methods section. The modules we are particularly interested in are the modules containing the top-ranking genes obtained based on feature importance scores. These top-ranking genes for each cell type are utilized as seed, under the assumption that genes with high rankings in state classification are most relevant to AD and thus serve as effective starting points for information dissemination. Figure 4c illustrates example modules using the top-ranking genes as seeds. Functional enrichment tests with KEGG pathways were performed for each identified module and the top three enriched pathways (all FDR < 0.05) for each of the fourteen modules are shown in Figure 4d. The comprehensive set of KEGG pathway enrichments for all modules can be found in the Supplementary files. As depicted in Figure 4d, certain pathways are recurrently detected across cell types; for instance, the "Thermogenesis" pathway is enriched in modules for Astrocyte, M. Fibro, Microglia, and Neuron. This finding aligns with recent studies suggesting the therapeutic potential of enhancing thermogenesis in AD[52]. Additionally, "PI3K-Akt signaling pathway" is enriched by modules detected for Arterial and Pericyte and "MAPK signaling pathway" is enriched by modules discovered for Neuron and OPC. Many researchers have suggested that "PI3K-Akt signaling pathway" and "MAPK signaling pathway" may be potential targets in preventing and treating AD[53-57]. Furthermore, other enriched pathways, strongly related to Alzheimer's disease, have garnered substantial literature support. For example, several studies report that "Oxidative phosphorylation" and "Neurotrophin signaling pathway" are related to AD[58,59]. As it happens, Neurotrophin signaling sits upstream of MAPK signaling—both are enriched in the Neuron module—and Neurotrophin modulated activation of MAPK and MAPK-ERK signaling has demonstrated increased neuronal survival and outgrowth, providing a potential target mechanism for therapeutic signaling agonists[60]. This suggests another potential benefit for ScAtt's independently derived network models: discovery of AD-related impacts that span multiple ontological groups or chains of multiple groups not always observable through standard linear or marginal associations and DEG pathway enrichment. Additionally, terms like "Alzheimer's disease" are significantly enriched in the module detected for Astrocyte (as detailed in the Supplementary files). In conclusion, these modules corroborate the earlier sections, consistently demonstrating that both the top-ranking genes and the modules they form are enriched for pathways relevant to AD, including the AD itself.

*Node centralities of the networks.* In our final analysis, we aimed to identify potential causal genes driving the AD state from ScAtt's network perspective. To achieve this, we conducted a topological analysis using three node centrality metrics: node degree, node closeness, and node PageRank value. These metrics were employed to rank nodes within the networks and assess the centrality of known AD-related genes in these networks. Conceptually, different centralities metrics quantify distinct types of biological importance. Node degree, the most basic centrality measure, is defined by the number of connections a node has with others. A gene with a higher degree suggests more interactions within the network. Node closeness centrality calculates the average shortest path between a node and all other nodes in the network, indicating how centrally located a gene is. Genes with high closeness centrality can rapidly influence or be influenced by the expression of other genes in the network. PageRank centrality, derived from a random walk through the network, measures the average time spent at a given node during all random walks. Analogous to the original PageRank for web pages, genes with high PageRank values can be considered as "popular" or influential within the network.

Supplementary Figure 2 illustrates the prioritization of known AD-related genes (from GWAS or Treat-AD) by these networks based on different node centralities and at various thresholds. We used different top K percentage thresholds, as indicated in the figure, to select a range of top-ranked genes based on node centralities for each cell type from networks constructed by different methods. Then, the total number of unique genes found across all fourteen cell types was aggregated. In terms of prioritizing known AD-related GWAS genes, ScAtt's performance was superior across all centralities, except for PageRank, where our network fell slightly below the Contrast Network at the top 5% threshold. However, at all other thresholds, ScAtt effectively identified the majority of AD-related genes as central to the network model. For known Treat-AD genes, our network consistently centralized more AD-related genes compared to other network models, regardless of the top K percentage threshold or centrality metric used. This analysis suggests that the topologically significant structures at the core of our networks are also of considerable importance in AD pathology.

**Methods**

**The scRNA-seq dataset used and data preprocessing**.

We evaluated the utility of ScAtt for single-cell transcriptome analysis using Single-cell RNA sequencing data (scRNAseq) from 95,186 single-nucleus transcriptomes from 17 hippocampus (9 AD cases and 8 controls). These samples were stratified into 14 distinct cell types annotated by previously published marker genes, which were collected and preprocessed by the Stanford/VA/NIA Aging Clinical Research Center (ACRC). The detailed preprocessing steps can be found in Yang et al[31]. Supplementary Figure 4a displays the dataset colored with individual samples, and Supplementary Figure 4b categorizes the same data by cell type. These figures illustrate the varied proportions of different cell types within the overall dataset and show a similar distribution of cell types across different individuals. The data were divided into training, validation, and test datasets based on the cohort samples, with 6 AD and 5 healthy controls in the training dataset, 1 AD and 1 healthy control in the validation dataset, and 2 AD and 2 healthy controls in the test dataset. As a result, the training dataset has 61,113 cells with 30,979 AD cells and 30,134 control cells, while the validation dataset has 12,088 cells with 5,320 AD cells and 6,768 control cells, and the test dataset has 21,985 cells with 10,387 AD cells and 11,598 control cells. There are 23,537 gene expression data as features per cell. The numbers of cells used for training, validation and test for each cell type are shown in Table 2.

**The architecture and parameters of our model**.

*The architecture of ScAtt*. The architecture of ScAtt is shown in Figure 1a. For a specific cell type, the cells of AD and Control cohorts are obtained, and each cell can be represented by the original expression of $g$ genes. After the shuffling procedure, a matrix $I \in \mathbb{R}^{c \times g}$ is formed as input, where $c$ is the number of cells and $g$ is the number of genes as features. Then, a linear layer converts the input matrix I into a matrix X with a lower dimensional representation, where $X \in \mathbb{R}^{c \times d_r}$. And then, the matrix X is delivered into the Transformer encoders. Each Transformer encoder has an attention system, feed-forward neural networks, and normalization steps as Figure 1a shows. The attention system is the core of the Transformer encoder architecture. It has $h$ parallel heads, so called multi-head attention. Each head has $N$ attention subsystems that perform the same task but process different inputs. After the linear layer, the encoder receives as input a representation matrix $X \in \mathbb{R}^{c \times d_r}$ for c cells, where the $d_r$ is the cell representation dimension and thus $N = c$. Here, we used $d_r = 512$ as the default setting for most of the cell types.

The input X goes through the attention system and first generates, for each cell, a query vector ($q_i$), a key vector ($k_i$), and a value vector ($v_i$), as followers:

$$Q_i = XW_i^Q, \quad K_i = XW_i^K, \quad V_i = XW_i^V$$

where $i$ is the index of the head, $W_i^Q$, $W_i^K$, and $W_i^V \in \mathbb{R}^{d_r \times d_k}$ are the learnable parameters and $d_k$ is the dimension of $Q_i$, $K_i$ and $V_i$. The value of $d_k$ is $d_r/h$, and the number of parallel heads $h = 8$ for the default setting for most cell types. $Q_i = \{q_1, q_2, \cdots, q_N\}$, $q_i \in \mathbb{R}^{1 \times d_k}$ is the queries matrix, $K_i = \{k_1, k_2, \cdots, k_N\}$, $k_i \in \mathbb{R}^{1 \times d_k}$ is the keys matrix, and $V_i = \{v_1, v_2, \cdots, v_N\}$, $v_i \in \mathbb{R}^{1 \times d_k}$ is the values matrix. Then, all $Q$, $K$, and $V$ are fed into the attention part of each parallel attention head, as Supplementary Figure 5 shows. The multi-head structure enables the model to explore multiple subspaces with various projections of the input data. The idea of having multiple heads on the Transformer is like having multiple filters on CNNs. Each head has $N$ attention subsystems that receive the same focus target representing as $K_i = \{k_1, k_2, \cdots, k_N\}$ as well as different cell inputs as $\{q_i\}$ and output an attention mask that connects all keys to a given query $q_i$. Mathematically, the operation performed by an attention head can be expressed as a matrix multiplication between all queries and keys, as seen below.

$$S_i = softmax\left(\frac{Q_i * K_i^T}{\sqrt{d_k}}\right)$$

where $S_i \in \mathbb{R}^{N \times N}$ is self-attention matrix and a softmax function converts the attention scores into probabilities. The interaction between $q_i$ and $k_i$ is a score of importance that reflects the interaction of features between the corresponding cells within a particular cell type. The attention mechanism allows parallel communication between different focused targets and input cell query, thus obtaining the global relations among features as attention mask between all the input cells. This system generates final attention mask $Att_m$ for the input $X$ of the encoder as below:

$$Z_i = S_i * V_i$$

$$Att_m = Concat(Z_1, Z_2, \cdots, Z_h) * W^O$$

where $Att_m = \{Att_{m,1}, Att_{m,2}, \cdots, Att_{m,N}\} \in \mathbb{R}^{N \times d_r}$ is a concatenation of each head, and $m \in [1, h]$, and $W^O \in R^{h \times d_k \times d_r}$ are the weights that will be learned during the model training procedure.

The output of multi-head attention then feeds into an Add & Normalize layer with a residual connection followed by a layer-normalization step, resulting in updated $X = \{x_1, x_2, \cdots, x_N\}$. Following that, each updated one $x_i$ goes through a feedforward neural network composed of linear transformations and a ReLU activation function. Finally, the residual input and feedforward results are fed into the final layer for the normalization step.

$$X = Norm(X + Att_m)$$

$$f_i = FeedForward(x_i) = max(0, x_i W_1 + b_1)W_2 + b_2$$

$$X = Norm(X + F)$$

where $F \in \mathbb{R}^{N \times d_k}$ is the feedforward matrix with elements of $f_i$, and the updated $X$ is the final output of the Transformer encoder. After a linear layer, the output of the encoder is converted into $Y \in \mathbb{R}^{N \times 2}$, and then a softmax function is performed to map the estimated output $y_i$ into a two-dimension vector with values ranging from (0, 1) as the probability of cell state classification. Here, the cross-entropy loss of binary classification defined as below is used.

$$L = -\frac{1}{N} \sum_{i=1}^{N} \sum_{j \in 0,1} t_{ij} \log y_{ij}$$

where $t_{ij}$ is the ground truth of $ith$ cell within a given cell type. $t_{i0} = 1$ represents the cell state of $ith$ cell is from control and $t_{i1} = 1$ represents the cell state of $ith$ cell is from AD. As $y_{i0}$ approaches 1, the $ith$ cell is more likely to be an AD cell. Our training process minimizes loss function $L$ by iteratively adjusting model's weights.

*Parameter settings and Parameter sensitivity analysis.* The model takes three hyperparameters: the parameter $d_r$ in the first linear layer (hidden dimension), the number of encoders $M$ and the number of multiple head attention $h$. We have discussed the relation between model size and the hyperparameters in the section of the stability and scalability of the model. The choice of these parameters depends on the sample size and the GPU memory size of the computer. For ScAtt, the default setting of $d_r$ for ten cell types with training sample size smaller than 7,500 is set to 512, the number of encoders $M$ is set to 3 and the number of multiple head attention $h$ is set to 8, while for training sample size larger than 7,500 including Oligo, Astrocyte, Pericyte and Capillary cell type, $d_r$, $M$ and $h$ are set to 16, 1 and 8, respectively. We analyze the model performance when using different $d_r$ (8, 16, 32, 64, 128, 256, and 512) for the cell types except for Oligo, Astrocyte, Pericyte and Capillary. For these four, we could only set $d_r$ to 8 or 16, as our model can only run with the $d_r$ up to 16 with a large sample size. Figure 2d shows the experimental results for with different hyperparameters on Venous and P. Fibro cell types and the remainder ten are shown in Supplementary Figure 6. From the experiment results, we can conclude that a higher $d_r$ leads to better performance of our model.

*Experimental setting for baseline methods*. In our experimental setup for baseline methods, the DNN was configured with a structure comprising three fully connected layers. For ResNet, we utilized the ResNet-34 architecture. These baseline models were trained for each cell type using the training dataset. To identify the optimal hyperparameters for the models, we employed a five-fold cross-validation strategy in conjunction with a grid search technique. Following the hyperparameter optimization, each model underwent retraining with the entire training dataset utilizing these optimal hyperparameters. The final model was established based on the criterion that there was no further improvement in performance on the validation dataset after 20 training iterations. Then, the final derived model was applied to the test dataset. The training process for the baseline models is illustrated in Supplementary Figure 7.

**Feature importance calculation**

One widely used method for both machine learning and deep learning models to assess the contribution of input to output in a trained model is weight-based feature importance. The weights are constantly modified to better fit the data during the training process. After the training is complete, it is natural to assume that the higher the absolute values of the weights, the more important a related feature is. Garson created one of the most well-known weighted-base approaches for measuring feature relevance in 1991[61], methodology that remains relevant today. Many approaches for evaluating feature importance were proposed based on Garson's and connection weights methods[62-65]. The basic idea of these methods is that all of the weights connecting a given feature to a specific result, even hidden layers, will contribute to evaluating feature importance. Inspired by this idea, here we propose a weighted-based method to evaluate the feature importance of the Transformer-based model. Given a training set $\{x_i, y_i\}_{i=1}^{N} \in \mathbb{R}^g$ with $N$ distinct cell samples for a cell type, each cell sample is described in $g$ gene features, and the model can learn the cell state classification between classes $y_i$ based on the input data $x_i$. As described above, the whole model architecture consists of a data process and input part, a linear layer, a Transformer component, and an output linear layer. For cell samples in a particular cell type with $g$ gene features, the classification output will be either AD or control. Define $W_{L1} \in \mathbb{R}^{g \times d_r}$, $W_T \in \mathbb{R}^{d_r \times d_r}$ and $W_{L2} \in \mathbb{R}^{d_r \times 2}$ to be the weight matrix of the first linear layer, Transformer encoder and the linear output layer, respectively. $W_T$ can be obtained by using the weight matrix of each layer within the Transformer encoder. Specifically, $W_T$ is obtained by Attention Rollout method[66], which recursively computes the attentions in each layer of the trained Transformer encoder given the input. Then, the importance matrix of gene features is calculated as:

$$IMP = |W_{L1}^0 W_T^0 W_{L2}^0 - W_{L1}^1 W_T^1 W_{L2}^1|$$

where the superscript of $W$ equals to 0 represents the weight connected to the output of control class and the superscript of $W$ equals 1 represents the weight connected to the output of AD class. $IMP \in \mathbb{R}^{g \times 1}$, and the importance of $i$th gene feature is $IMP_i$, which indicates the importance of the gene in classifying the cell state of a particular cell type into AD and Control. The IMP can approximate how input genes contribute to a predicted output.

**Network construction and Module detection**

*Constructing regulatory network vis ScAtt*. Our model leverages the relationships among all cells within a cell type, as identified by the Transformer, to classify cell state into AD and Control categories. After obtaining the well-trained model, each gene feature will have a vector embedding $v_i$ in our model, derived by multiplying all the weights of the model's parameters. The multiplication results illustrate the role of the gene feature in identifying the cell state. To assess whether pairs of genes play similar roles in distinguishing cell status, we employed cosine similarity. Additionally, we considered the feature importance score of each gene as an indicator of its magnitude of contribution to the cell state classification. Then the relation score of two genes $(i, j)$ can be defined as:

$$Score(i, j) = cosine(v_i, v_j) * IMP_i * IMP_j \geq threshold_{weight}$$

Based on the score, the gene pairs whose score is greater than $threshold_{weight}$ value are selected to construct a network for a cell type.

*Constructing benchmark networks*. For the comparison networks, we used the Pearson correlation of gene expressions to calculate the co-expression between two genes. Hence the relation score of two genes $(i, j)$ can be defined as

$$Score(i, j) = |Pearson(e_i, e_j)| \geq threshold_{weight}$$

Where $e_i$ and $e_j$ are vectors of the expression values of gene $i$ and gene $j$ on either Control or AD samples after normalization. The gene pairs with high weight are assumed to be highly reliable and then selected to construct the AD and Control Networks. For constructing the Contrast Network, we first calculate the Pearson correlation of gene pairs for AD and Control, then we qualify the difference between AD and Control by the following equation.

$$Score_{Contrast}(i, j) = |Score_{AD}(i, j) - Score_{Control}(i, j)| \geq threshold_{weight}$$

*The weight thresholds selection*. We retain the significate correlations by setting the weight threshold to construct the final network. Following the previous study, in order to adjust for technical factors when constructing networks, an adaptive rather than fixed correlation threshold is used[67]. Specifically, the inferred networks were constructed by retaining the top $1/N$ relations where $N$ is the number of genes. Using this relative threshold prevents technical factors from producing artificial differences when comparing different networks.

*The module detection method.* There have been extensive studies for identifying modules from graphs or networks, leading to the development of various module detection methods and clustering algorithms. These techniques have been successfully applied to discover functional modules or protein complexes from various types of biological networks such as protein interaction networks. However, the selection of an appropriate method is contingent upon the specific type of biological network being analyzed. In our case, the networks of AD-related genes are relatively dense, and we have seed genes that we want to examine. So we devise a seed-based module detection method that uses alias sampling strategy[68] based on the edge weights to select the potential addable node and uses cohesiveness[69] to determine whether to add the selected node to the current module. The cohesiveness of a module $V$ is given as:

$$f(V) = \frac{w^{in}(V)}{w^{in}(V) + w^{bound}(V) + p|V|}$$

Where $w^{in}(V)$ denotes the total weight of edges contained in a module $V$, $w^{bound}(V)$ denotes the total weight of edges that connect the module $V$ with the rest of the network, and $p(V)$ is a penalty term whose purpose is to model the uncertainty in the data by assuming the existence of yet undiscovered interactions in the network. Given seed node *s*, *s* is the only member of an initial module *V*. By using alias sampling strategy, we could find potential additive node $u$, and if $f(V+u) > f(V)$, then $u$ can be included in *V* otherwise end the extension process. This module detection method allows the detection of dense modules that are well separated from the rest of the networks.

**Conclusion.**

It is still a fundamental challenge to explore cellular heterogeneity in scRNA-seq data. ScAtt's architectural innovation lies in its ability to integrate both local and global information across all features across cells within a specific cell type, thereby accurately determining a cell's disease status through capturing the complex linear and nonlinear patterns in the scRNA-seq data. While accurate predictions are important, it is equally important to be able to mine the features driving prediction, which provides valuable information about the model itself. ScAtt's Attention-based Transformer architecture allows for the extraction of gene features that drive disease state classifications. It provides feature embeddings, and the gene-gene networks constructed from these embeddings show a correlation with AD disease status. ScAtt acknowledges the heterogeneity among different cell types, uncovering cell-type-specific AD-related genes and constructing corresponding gene regulatory networks. Taking it one step further, we identified cell-type-specific modules, thereby enhancing our understanding of disease-associated pathways in AD that are unique to particular cell types. Having compared the ScAtt derived cell type modules to AD-related pathways, we see a very encouraging enrichment of functional information at the center of our model's prediction accuracy. As our understanding of the heterogeneity of AD phenotypes has grown in recent years, it has become apparent that broad disruption across many cell types is likely, and ScAtt offers a unique tool to explore these varying disruptions.

However, our method does have limitations. It is prone to achieve better classification results with medium to large cell-type data sets compared to relatively small groups (e.g.,≤124 cells as T cell and M. Fibro), as it is designed to leverage the global cell type features amongst many cells. Additionally, very large groups are computationally limited (e.g., ≥7500 cells as Capillary, Pericyte, Astrocyte and Oligo). While the model outperforms other baseline methods, the Transformer can only process an input dimension of up to 16 on a single GPU with 16GB memory. Future efforts will focus on developing a more efficient model architecture capable of handling large, high-dimensional datasets, even with limited computational resources. This is particularly relevant as our analysis indicates that higher input dimensions can lead to enhanced performance. Beyond AD, ScAtt's gene-feature aware classification has considerable potential to impact various complex diseases—through diagnosis/stratification, discovery of gene targets, and implication of cell types—hopefully leading to more personalized disease treatment, and improved study of disease heterogeneity.

**Data Availability and Code Availability**

The scRNA-Seq data is available at https://www.ncbi.nlm.nih.gov/geo/query/acc.cgi?acc=GSE163577. And the code can be accessed at https://github.com/circustata/ScAtt.

**Acknowledgments**

This research was supported by NIH/NIA award AG066206 (ZH).

**Author Contributions**

X.L. and Z.H. developed the concepts for the manuscript and proposed the method. X.L., R.R.B., P.K.G., and Z.H. designed the analyses and applications and discussed results. X.L., Z.H., R.R.B. and F.M.L prepared the manuscript and contributed to editing the paper.

**Competing Interests**

The authors declare no competing interests.

**Figure 1. Overview of the ScAtt workflow.** The three panels show a) The illustration of ScAtt's architecture. There are mainly three parts: data processing, model framework, and model training. b) Obtaining cell-type-specific AD-related genes from the model. The parameters of the trained model for each cell type were used to measure feature importance scores and then use feature importance scores to identify AD-related genes. c) Cell-type-specific gene-gene network construction and further analysis of the constructed networks. The feature embeddings provided by the model were used to calculate gene-gene correlations and by which we construct AD-related gene-gene networks. After constructing the networks, module detection and enrichment analysis were performed.

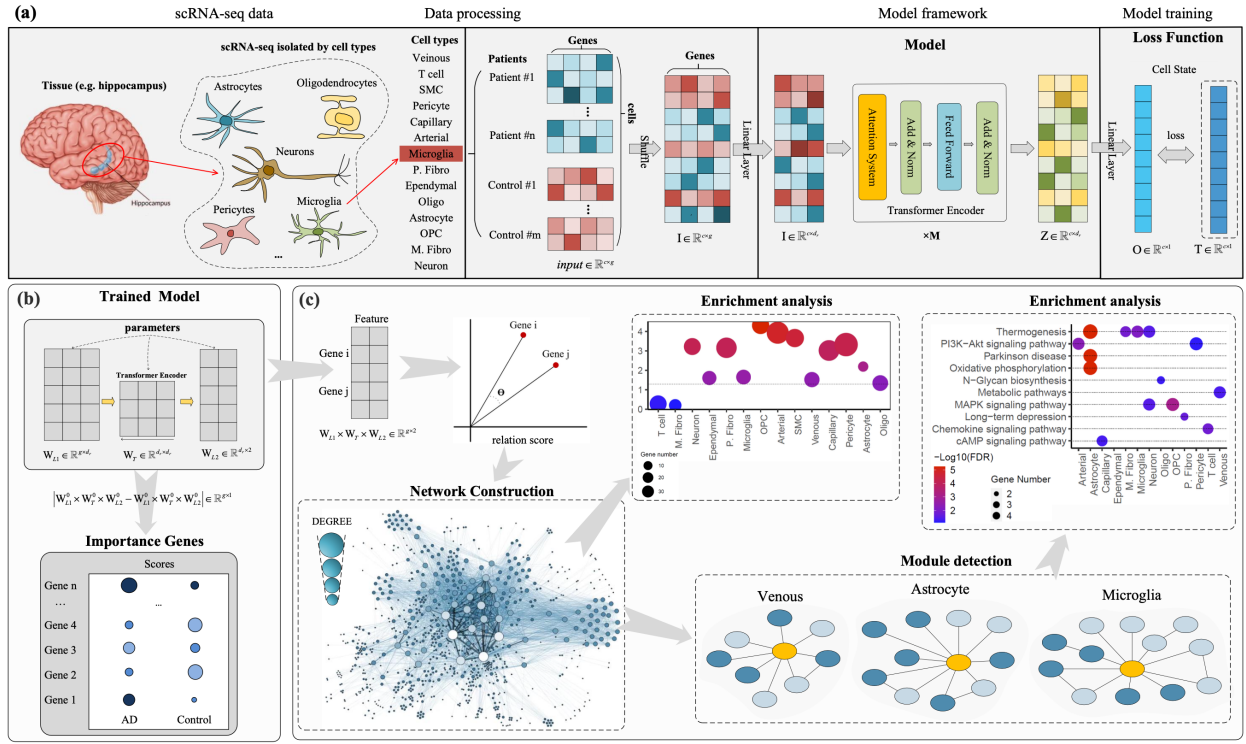

**Figure 2. The performance of model prediction and feature learning.** a) The different training strategies between machine learning models and our method. b) The AUC performance of cell state predicting for all methods, and the bottom right is the cell distribution in all training data. c) The performance of cross-cell type. Each row and column represent different cell types, and the value is the AUC performance on the corresponding column cell type with the model trained on the row cell type. The best performance for a row is marked with a red box. d) The effects of hyperparameters on model size and performance. The X-axis indicates the number of encoders, and the y-axis indicates the input dimension of the model. The model size is defined by the size of the model's parameters.

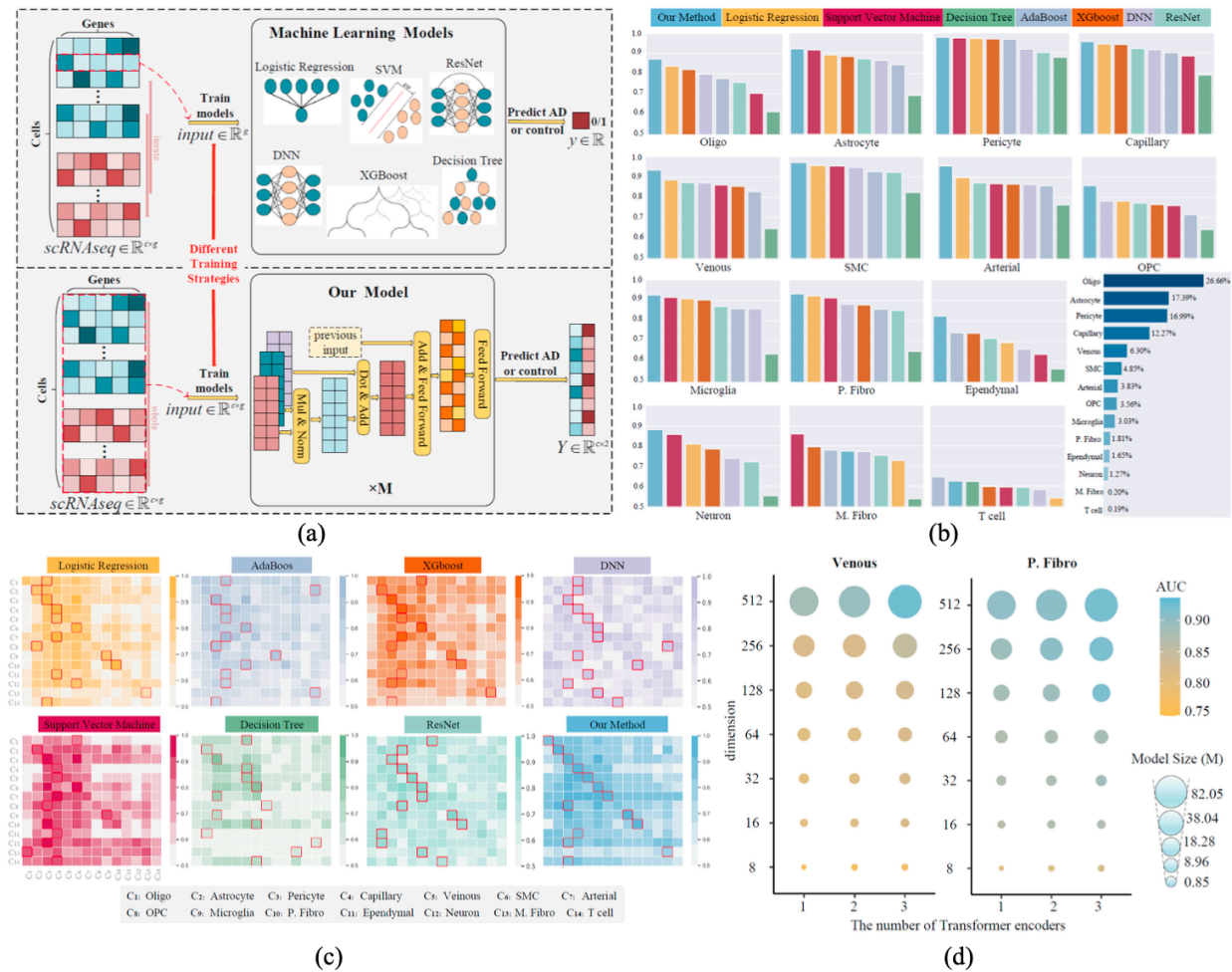

**Figure 3. The ability of discovering AD-related genes.** We compared the union of genes found for each cell type. a) Comparison of our method and seven machine learning algorithms for identifying known GWAS and Treat-AD genes when selecting different top K genes based on feature importance. The x-axis is the top-K threshold. b) Comparison of ScAtt and Seurat for identifying known GWAS and Treat-AD genes when using a different p-value of Seurat to select genes. c) Venn diagram illustrating the overlap between known GWAS or Treat-AD genes with ScAtt and Seurat when the p-value is equal to 0.05. d) The intersection heatmap of found genes between 14 cell types. The top two are based on known GWAS genes, and the bottom two are based on known Treat-AD genes.

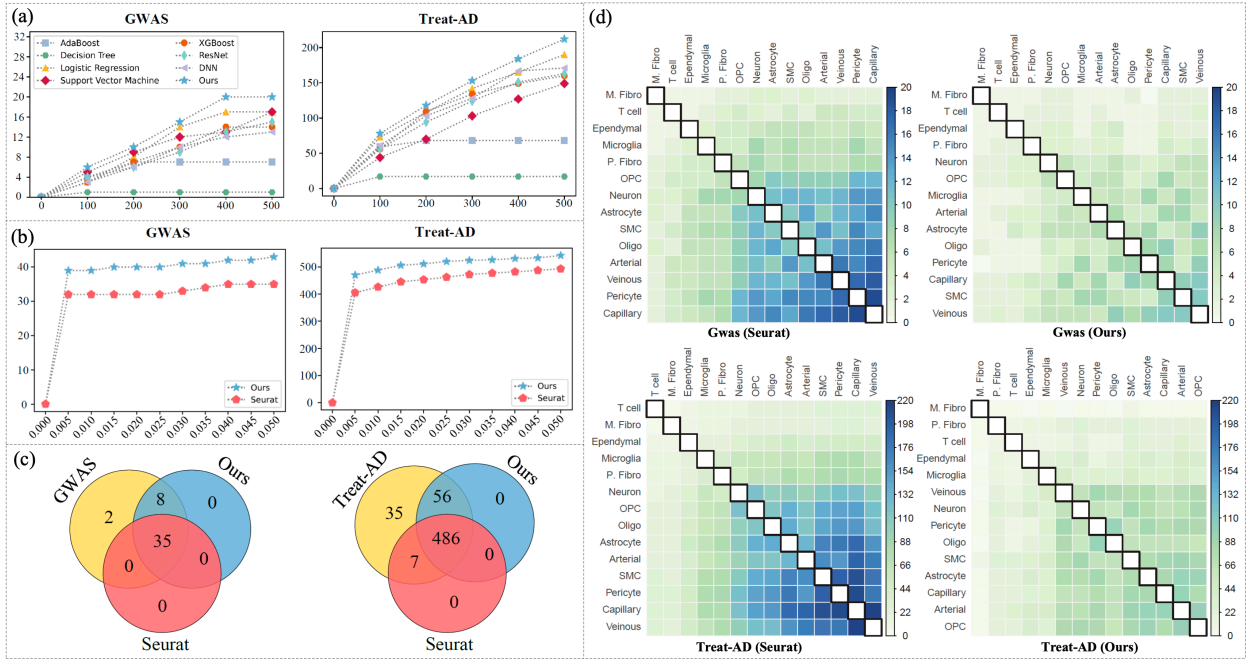

**Figure 4. Networks construction and modules detection.** a) The illustration of constructing networks. The upper one is for constructing AD, Control and Contrast Networks. AD Network and Control Network are constructed by using the Pearson's correlation coefficient (PCC) score of the gene expressions of AD patients and control cohorts, respectively. Contrast Network is constructed by contrasting the difference between AD and Control network. The lower one is for constructing our network, and the construction details are in the method. b) The "Alzheimer's disease" pathway (hsa05010) enrichment results for the four networks. The x-axis represents different cell types and the y-axis represents the Enrichment Ratio (ER), ER = (# overlapped gene / # total gene in the pathway) / (# genes in network / # total gene). The values on the top of the bars indicate the corresponding p-value for our networks. c) The example modules detected for four cell types, and the seed genes are marked with yellow color. d) The top three KEGG enrichment results for the modules detected for fourteen cell types.

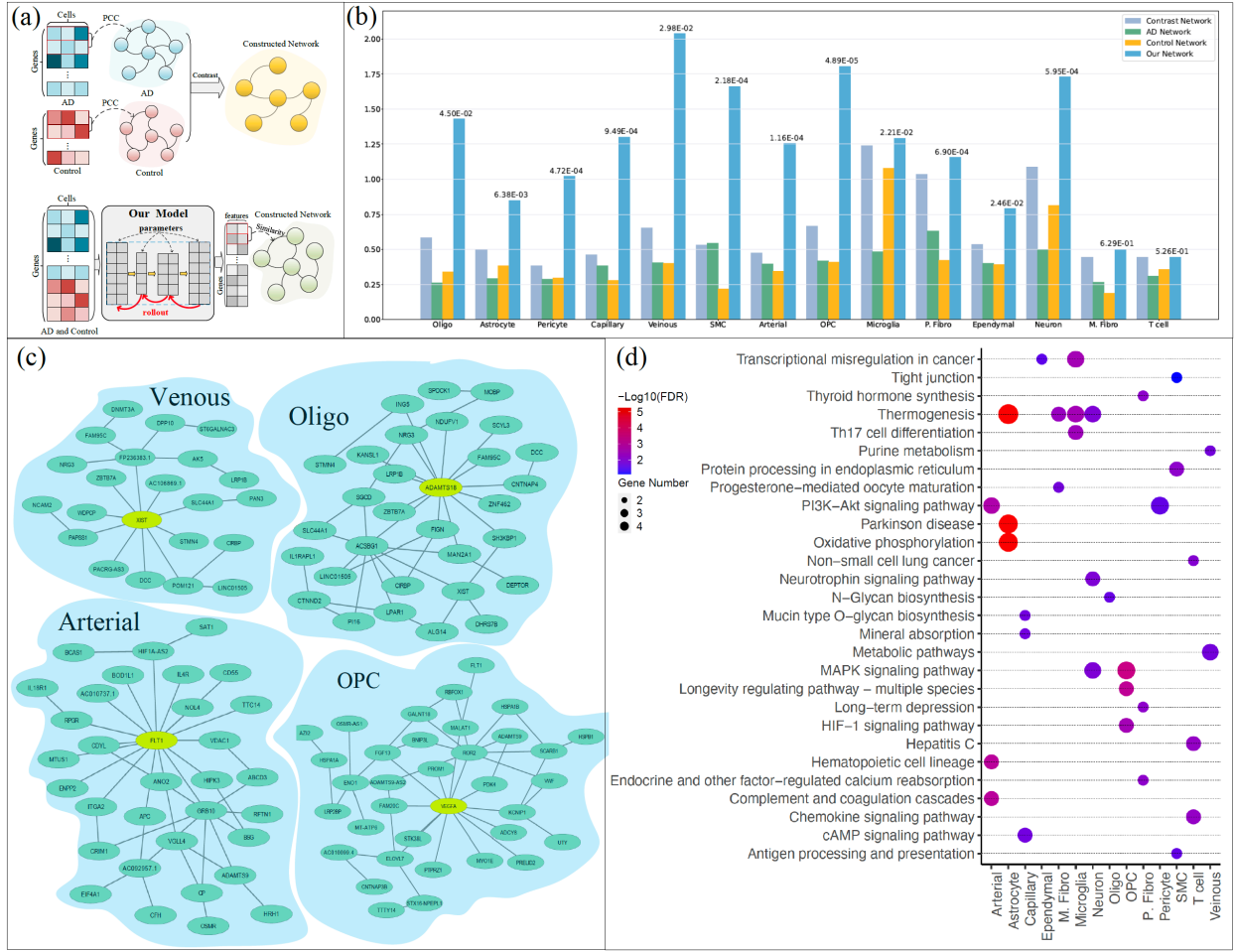

**Table 1. The top five ranked genes based on feature importance for each cell type.**

| Cell Type | Top 1 | Top 2 | Top 3 | Top 4 | Top 5 |
|---|---|---|---|---|---|
| Arterial | FLT1 | AC092957.1 | ADAMTS9 | ABCD3 | CP |
| Astrocyte | LINC00278 | DGKB | MT-ATP6 | TMTC1 | MT-ND2 |
| Capillary | ADAMTS9 | PDE10A | BACE2 | AC092957.1 | CP |
| Ependymal | TTTY14 | UTY | AC092957.1 | AC019330.1 | PDE4DIP |
| M. Fibro | PCDHA1 | MAPK8IP1 | PRR29-AS1 | TRPC3 | CNTN4-AS1 |
| Microglia | ACSL1 | SLC11A1 | XIST | PDE3B | HSPB1 |
| Neuron | MT-ATP6 | XKR6 | MT-ND4 | NEAT1 | TTTY14 |
| Oligo | ADAMTS18 | XIST | MAN2A1 | CIRBP | LINC01505 |
| OPC | VEGFA | ADAMTS9-AS2 | BNIP3L | RBFOX1 | GALNT18 |
| P. Fibro | XIST | CEMIP | ADAMTS12 | AC093772.1 | FLRT2 |
| Pericyte | XIST | SLC6A1-AS1 | APBB2 | SLC20A2 | ATP1A2 |
| SMC | ADAMTS9 | PCBP3 | IGFBP4 | DNAJB1 | CEBPD |
| T cell | NEAT1 | ADAMTS9 | MBD5 | HSPB1 | UTY |
| Venous | XIST | SLC39A10 | GPCPD1 | APBB2 | ABCG2 |

**Table 2. The number of cells in the training, validation, and test datasets for 14 cell types.**

| Cell type | Training cell number | Validation cell number | Test cell number |
|---|---|---|---|
| Oligo | 16,293 | 2,743 | 3,880 |
| Astrocyte | 10,627 | 2,287 | 3,840 |
| Pericyte | 10,383 | 1,610 | 3,791 |
| Capillary | 7,500 | 1,847 | 3,111 |
| Venous | 3,852 | 1,011 | 1,897 |
| SMC | 2,961 | 899 | 1,570 |
| Arterial | 2,338 | 606 | 1,433 |
| OPC | 2,177 | 308 | 624 |
| Microglia | 1,854 | 244 | 428 |
| P. Fibro | 1,104 | 209 | 641 |
| Ependymal | 1,008 | 183 | 289 |
| Neuron | 778 | 50 | 270 |
| M. Fibro | 124 | 60 | 117 |
| T cell | 114 | 31 | 94 |

**Supplementary Material**

**The results of cell state classification at the patient level**.

Prediction of patient phenotypes from scRNA-seq data is difficult for a number of reasons, including the varying number of cells in the expression profile and the limited number of available patients. After obtaining cell-level state classification, we aggregated these results to predict patient-level state classification. Supplementary Figure 3 illustrates the results of this patient-level state classification. There are four individuals in our test data, 2 AD patients and 2 healthy controls. Each individual contributes a differing number of cells across various cell types. For AD status classification, we employed a simple majority rule (>50%), considering the proportion of cells per cell type and per individual. Out of 56 test groups, 53 were accurately predicted, except for one group each in the Ependymal, T cell, and Capillary cell types, resulting in an overall accuracy of 94.6%. We also introduced the "single-cell type score," a metric representing the average accuracy of predictions across all individuals. This metric assesses the diagnostic significance of different cell types for Alzheimer's disease. ScAtt demonstrated its ability to accurately diagnose Alzheimer's disease using data from most single cell types. The combined data set (ALL) shows an average of 83.5% of cells correctly classified, indicating high confidence in our patient-level disease state predictions. Moreover, our method is also applicable to diagnosing Alzheimer's disease using data from all cell types. The figure shows that ScAtt correctly classifies Alzheimer's patients using all cell type data, as shown in the column for ALL. Interestingly, specific cell types, such as SMC, P. Fibro, or Arterial, were found to diagnose Alzheimer's disease more accurately on their own than using the combined data set, hinting at the potential of these cell types in future diagnostic processes. However, given the limited set of individuals in the test set, a much more robust cohort will be needed to generate definitive classification accuracy. Another considerable challenge in using ScAtt (and scRNA-seq in general) for diagnostics is obtaining brain tissue from living participants. Nevertheless, given ScAtt's ability to capture gene feature information, future research could explore its functional overlap with circulating cell types (e.g., from blood samples), potentially paving the way for a new, powerful diagnostic tool.

**Supplementary Table 1. The detailed results of the cell state identification for 14 cell types.**

| Method | AUC | ACC[a] | PPV[b] | Sensitivity | F-score | Specificity | NPV[c] |
|---|---|---|---|---|---|---|---|
| | | | Venous | | | | |
| Logistic Regression | 0.886 | 0.763 | 0.855 | 0.583 | 0.693 | 0.583 | 0.720 |
| Support Vector Machine | 0.861 | 0.756 | 0.797 | 0.631 | 0.705 | 0.631 | 0.733 |
| Decision Tree | 0.645 | 0.541 | 1.000 | 0.003 | 0.007 | 0.003 | 0.541 |
| AdaBoost | 0.828 | 0.724 | 0.788 | 0.549 | 0.647 | 0.549 | 0.694 |
| XGboost | 0.854 | 0.746 | 0.816 | 0.580 | 0.678 | 0.580 | 0.713 |
| ResNet | 0.873 | 0.746 | 0.733 | 0.721 | 0.727 | 0.768 | 0.757 |
| DNN | 0.872 | 0.787 | 0.838 | 0.704 | 0.765 | 0.704 | 0.751 |
| ScAtt | **0.935** | 0.849 | 0.877 | 0.781 | 0.826 | 0.781 | 0.829 |
| | | | SMC | | | | |
| Logistic Regression | 0.958 | 0.892 | 0.919 | 0.838 | 0.876 | 0.838 | 0.874 |
| Support Vector Machine | 0.955 | 0.894 | 0.895 | 0.870 | 0.882 | 0.870 | 0.894 |
| Decision Tree | 0.824 | 0.762 | 0.705 | 0.819 | 0.758 | 0.819 | 0.826 |
| AdaBoost | 0.927 | 0.854 | 0.856 | 0.815 | 0.835 | 0.815 | 0.852 |
| XGboost | 0.923 | 0.846 | 0.819 | 0.849 | 0.834 | 0.849 | 0.870 |
| ResNet | 0.923 | 0.800 | 0.786 | 0.815 | 0.800 | 0.786 | 0.815 |
| DNN | 0.948 | 0.863 | 0.893 | 0.794 | 0.841 | 0.794 | 0.843 |
| ScAtt | **0.972** | 0.923 | 0.942 | 0.885 | 0.913 | 0.885 | 0.909 |
| | | | Arterial | | | | |
| Logistic Regression | 0.898 | 0.746 | 0.915 | 0.534 | 0.674 | 0.534 | 0.678 |
| Support Vector Machine | 0.867 | 0.749 | 0.871 | 0.575 | 0.693 | 0.575 | 0.690 |
| Decision Tree | 0.763 | 0.620 | 0.732 | 0.360 | 0.482 | 0.360 | 0.584 |
| AdaBoost | 0.857 | 0.763 | 0.830 | 0.652 | 0.730 | 0.652 | 0.720 |
| XGboost | 0.865 | 0.760 | 0.830 | 0.644 | 0.726 | 0.644 | 0.716 |
| ResNet | 0.872 | 0.662 | 0.617 | 0.685 | 0.649 | 0.644 | 0.709 |
| DNN | 0.864 | 0.710 | 0.884 | 0.475 | 0.618 | 0.475 | 0.648 |
| ScAtt | **0.956** | 0.872 | 0.922 | 0.807 | 0.861 | 0.807 | 0.833 |
| | | | P. Fibro | | | | |
| Logistic Regression | 0.922 | 0.716 | 0.931 | 0.466 | 0.621 | 0.466 | 0.644 |
| Support Vector Machine | 0.912 | 0.727 | 0.914 | 0.500 | 0.646 | 0.500 | 0.657 |
| Decision Tree | 0.652 | 0.502 | 1.000 | 0.003 | 0.006 | 0.003 | 0.502 |
| AdaBoost | 0.856 | 0.657 | 0.917 | 0.344 | 0.500 | 0.344 | 0.597 |
| XGboost | 0.877 | 0.672 | 0.958 | 0.359 | 0.523 | 0.359 | 0.607 |
| ResNet | 0.850 | 0.675 | 0.753 | 0.618 | 0.679 | 0.746 | 0.609 |
| DNN | 0.880 | 0.727 | 0.857 | 0.544 | 0.665 | 0.544 | 0.667 |
| ScAtt | **0.932** | 0.856 | 0.835 | 0.887 | 0.861 | 0.887 | 0.880 |
| | | | Ependymal | | | | |
| Logistic Regression | 0.694 | 0.637 | 0.451 | 0.327 | 0.379 | 0.327 | 0.697 |
| Support Vector Machine | 0.637 | 0.633 | 0.465 | 0.541 | 0.500 | 0.541 | 0.743 |
| Decision Tree | 0.565 | 0.626 | 0.444 | 0.408 | 0.426 | 0.408 | 0.709 |
| AdaBoost | 0.740 | 0.692 | 0.622 | 0.235 | 0.341 | 0.235 | 0.702 |
| XGboost | 0.738 | 0.692 | 0.615 | 0.245 | 0.350 | 0.245 | 0.704 |
| ResNet | 0.712 | 0.700 | 0.761 | 0.635 | 0.692 | 0.773 | 0.652 |
| DNN | 0.662 | 0.629 | 0.689 | 0.635 | 0.661 | 0.635 | 0.562 |
| ScAtt | **0.823** | 0.709 | 0.547 | 0.827 | 0.659 | 0.827 | 0.879 |
| | | | Microglia | | | | |
| Logistic Regression | 0.909 | 0.783 | 0.947 | 0.656 | 0.775 | 0.656 | 0.676 |
| Support Vector Machine | 0.916 | 0.745 | 0.953 | 0.582 | 0.723 | 0.582 | 0.634 |
| Decision Tree | 0.639 | 0.636 | 0.718 | 0.594 | 0.650 | 0.594 | 0.562 |
| AdaBoost | 0.860 | 0.738 | 0.847 | 0.660 | 0.742 | 0.660 | 0.651 |
| XGboost | 0.903 | 0.794 | 0.915 | 0.705 | 0.796 | 0.705 | 0.700 |
| ResNet | 0.872 | 0.712 | 0.713 | 0.692 | 0.711 | 0.701 | 0.714 |
| DNN | 0.859 | 0.619 | 0.685 | 0.615 | 0.648 | 0.615 | 0.550 |
| ScAtt | **0.928** | 0.813 | 0.956 | 0.705 | 0.811 | 0.705 | 0.710 |
| OPC | | | | | | | |
| Logistic Regression | 0.780 | 0.720 | 0.676 | 0.627 | 0.651 | 0.627 | 0.747 |
| Support Vector Machine | 0.759 | 0.694 | 0.639 | 0.612 | 0.625 | 0.612 | 0.731 |
| Decision Tree | 0.640 | 0.627 | 0.536 | 0.769 | 0.632 | 0.769 | 0.761 |
| AdaBoost | 0.714 | 0.660 | 0.575 | 0.712 | 0.636 | 0.712 | 0.752 |
| XGboost | 0.764 | 0.692 | 0.622 | 0.665 | 0.643 | 0.665 | 0.749 |
| ResNet | 0.771 | 0.700 | 0.686 | 0.706 | 0.696 | 0.694 | 0.714 |
| DNN | 0.781 | 0.607 | 0.521 | 0.727 | 0.607 | 0.727 | 0.728 |
| ScAtt | **0.858** | 0.720 | 0.615 | 0.877 | 0.723 | 0.877 | 0.874 |
| Neuron | | | | | | | |
| Logistic Regression | 0.809 | 0.774 | 0.684 | 0.600 | 0.639 | 0.600 | 0.812 |
| Support Vector Machine | 0.857 | 0.781 | 0.682 | 0.644 | 0.663 | 0.644 | 0.827 |
| Decision Tree | 0.551 | 0.619 | 0.420 | 0.378 | 0.398 | 0.378 | 0.704 |
| AdaBoost | 0.721 | 0.711 | 0.567 | 0.567 | 0.567 | 0.567 | 0.783 |
| XGboost | 0.784 | 0.733 | 0.594 | 0.633 | 0.613 | 0.633 | 0.810 |
| ResNet | 0.721 | 0.714 | 0.735 | 0.694 | 0.714 | 0.735 | 0.694 |
| DNN | 0.737 | 0.695 | 0.796 | 0.455 | 0.579 | 0.455 | 0.660 |
| ScAtt | **0.881** | 0.752 | 0.595 | 0.800 | 0.682 | 0.800 | 0.879 |
| M. Fibro | | | | | | | |
| Logistic Regression | 0.727 | 0.675 | 0.588 | 0.638 | 0.612 | 0.638 | 0.742 |
| Support Vector Machine | **0.859** | 0.718 | 0.613 | 0.809 | 0.697 | 0.809 | 0.836 |
| Decision Tree | 0.537 | 0.496 | 0.427 | 0.745 | 0.543 | 0.745 | 0.657 |
| AdaBoost | 0.779 | 0.650 | 0.544 | 0.787 | 0.643 | 0.787 | 0.796 |
| XGboost | 0.795 | 0.718 | 0.613 | 0.809 | 0.697 | 0.809 | 0.836 |
| ResNet | 0.753 | 0.662 | 0.675 | 0.675 | 0.675 | 0.649 | 0.649 |
| DNN | 0.772 | 0.619 | 0.563 | 0.681 | 0.616 | 0.568 | 0.685 |
| ScAtt | 0.774 | 0.701 | 0.607 | 0.723 | 0.660 | 0.723 | 0.787 |
| T cell | | | | | | | |
| Logistic Regression | 0.540 | 0.521 | 0.500 | 0.778 | 0.609 | 0.778 | 0.583 |
| Support Vector Machine | 0.594 | 0.553 | 0.526 | 0.667 | 0.588 | 0.667 | 0.595 |
| Decision Tree | 0.623 | 0.606 | 0.583 | 0.622 | 0.602 | 0.622 | 0.630 |
| AdaBoost | **0.645** | 0.628 | 0.581 | 0.800 | 0.673 | 0.800 | 0.719 |
| XGboost | 0.597 | 0.564 | 0.530 | 0.778 | 0.631 | 0.778 | 0.643 |
| ResNet | 0.593 | 0.649 | 0.667 | 0.533 | 0.593 | 0.533 | 0.638 |
| DNN | 0.582 | 0.574 | 0.407 | 0.561 | 0.472 | 0.561 | 0.721 |
| ScAtt | 0.624 | 0.574 | 0.558 | 0.533 | 0.545 | 0.533 | 0.588 |
| Pericyte | | | | | | | |
| Logistic Regression | 0.971 | 0.848 | 0.962 | 0.695 | 0.807 | 0.695 | 0.792 |
| Support Vector Machine | 0.973 | 0.865 | 0.967 | 0.728 | 0.831 | 0.728 | 0.811 |
| Decision Tree | 0.880 | 0.765 | 0.908 | 0.540 | 0.677 | 0.540 | 0.712 |
| AdaBoost | 0.967 | 0.838 | 0.959 | 0.673 | 0.791 | 0.673 | 0.781 |
| XGboost | 0.968 | 0.858 | 0.949 | 0.726 | 0.823 | 0.726 | 0.808 |
| ResNet | 0.901 | 0.853 | 0.828 | 0.874 | 0.850 | 0.834 | 0.879 |
| DNN | 0.918 | 0.853 | 0.797 | 0.908 | 0.849 | 0.908 | 0.913 |
| ScAtt | **0.976** | 0.923 | 0.917 | 0.913 | 0.915 | 0.913 | 0.928 |
| Capillary | | | | | | | |
| Logistic Regression | 0.943 | 0.726 | 0.975 | 0.519 | 0.678 | 0.519 | 0.621 |
| Support Vector Machine | 0.887 | 0.737 | 0.906 | 0.586 | 0.712 | 0.586 | 0.642 |
| Decision Tree | 0.792 | 0.596 | 0.966 | 0.282 | 0.436 | 0.282 | 0.525 |
| AdaBoost | 0.900 | 0.724 | 0.931 | 0.543 | 0.686 | 0.543 | 0.625 |
| XGboost | 0.941 | 0.778 | 0.956 | 0.629 | 0.758 | 0.629 | 0.676 |
| ResNet | 0.921 | 0.760 | 0.811 | 0.720 | 0.762 | 0.806 | 0.714 |
| DNN | 0.915 | 0.723 | 0.955 | 0.525 | 0.678 | 0.525 | 0.621 |
| ScAtt | **0.955** | 0.833 | 0.965 | 0.725 | 0.828 | 0.725 | 0.738 |
| | | | Oligo | | | | |
| Logistic Regression | 0.838 | 0.760 | 0.763 | 0.796 | 0.779 | 0.796 | 0.757 |
| Support Vector Machine | 0.702 | 0.653 | 0.650 | 0.753 | 0.698 | 0.753 | 0.658 |
| Decision Tree | 0.610 | 0.561 | 0.552 | 0.921 | 0.690 | 0.921 | 0.630 |
| AdaBoost | 0.776 | 0.695 | 0.657 | 0.894 | 0.757 | 0.894 | 0.796 |
| XGboost | 0.821 | 0.730 | 0.685 | 0.911 | 0.782 | 0.911 | 0.838 |
| ResNet | 0.756 | 0.745 | 0.746 | 0.784 | 0.765 | 0.702 | 0.745 |
| DNN | 0.798 | 0.707 | 0.684 | 0.751 | 0.716 | 0.751 | 0.734 |
| ScAtt | **0.873** | 0.788 | 0.774 | 0.850 | 0.810 | 0.850 | 0.809 |
| | | | Astrocyte | | | | |
| Logistic Regression | 0.891 | 0.805 | 0.811 | 0.640 | 0.715 | 0.640 | 0.802 |
| Support Vector Machine | 0.913 | 0.820 | 0.853 | 0.641 | 0.732 | 0.641 | 0.806 |
| Decision Tree | 0.690 | 0.653 | 0.539 | 0.667 | 0.596 | 0.667 | 0.757 |
| AdaBoost | 0.842 | 0.767 | 0.730 | 0.625 | 0.673 | 0.625 | 0.786 |
| XGboost | 0.884 | 0.798 | 0.779 | 0.661 | 0.715 | 0.661 | 0.807 |
| ResNet | 0.872 | 0.663 | 0.699 | 0.642 | 0.669 | 0.688 | 0.630 |
| DNN | 0.864 | 0.710 | 0.884 | 0.475 | 0.618 | 0.475 | 0.648 |
| ScAtt | **0.919** | 0.837 | 0.756 | 0.849 | 0.799 | 0.849 | 0.898 |

[a]ACC stands for accuracy. [b]PPV stands for positive predictive value. [c]NPV stands for negative predictive value.

**Supplementary Figure 1. Comparison results between Seurat and our method using adjusted p-value as threshold.** a) Comparison of ScAtt and Seurat for identifying known GWAS and Treat-AD genes when using a different adjusted p-value of Seurat to select genes. b) Venn diagram illustrating the overlap between known GWAS or Treat-AD genes with ScAtt and Seurat when the adjusted p-value is equal to 0.05.

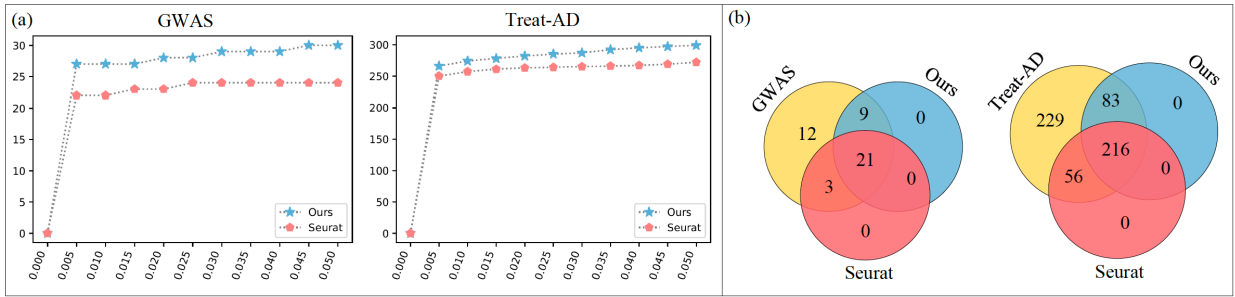

1  **Supplementary Figure 2: The number of genes discovered by competing networks based on deferent node centralities by
2  using different top k thresholds.** (a) shows the numbers of AD-related genes found based on node degree with different top K
3  percentage ranking cutoffs. (b) shows the numbers of AD-related genes found based on node closeness centrality with different top
4  K percentage ranking cutoffs. (c) shows the numbers of AD-related genes found based on node PageRank value with different top
5  K percentage ranking cutoffs. The results of the number of GWAS genes discovered are shown on the left, and the results of the
6  number of Treat-AD genes discovered are shown on the right.

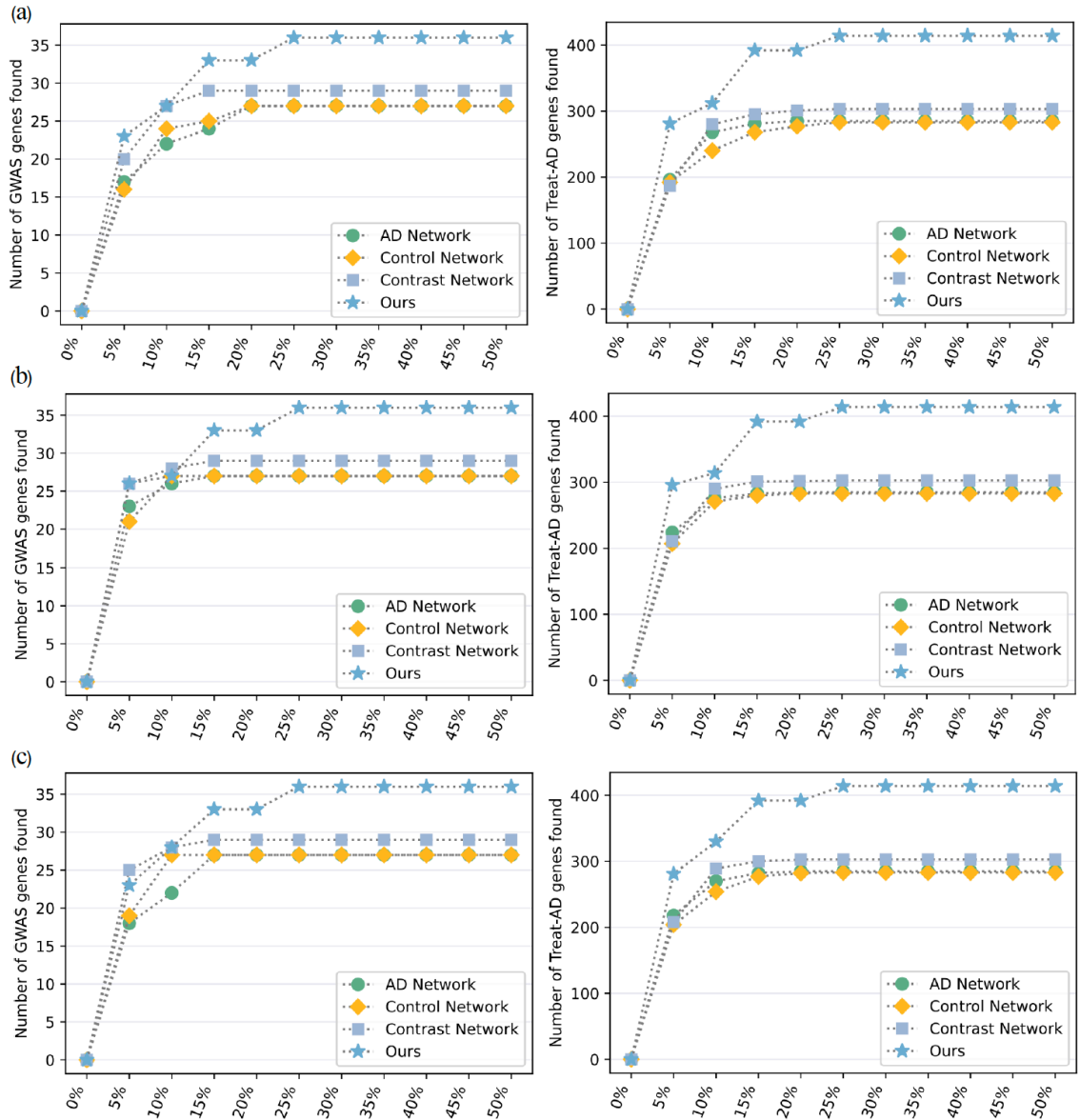

**Supplementary Figure 3. The classification results at the patient level.** It shows the predicted results of each cell type. AD and Control denote AD patients and control individuals, respectively. The number on the right of the slash is the cell amount, and the number on the left indicates the proportion of cells belonging to AD patients predicted by our model. For example, the value 0.79/817 (the top-left data corresponding to AD1-Venous) indicates that our model predicts that 79% (645 cells) of the 817 Venous cells belong to AD patients. When more than 50% of cells are predicted to be AD cells, the data box is marked in red, indicating that the patient is predicted as an AD patient. The box colored in blue is considered as the control group by our model. We defined the "Single Cell Type Score" metric, which is the average of predicted proportion for all the ADs and Controls, to evaluate the importance of different cell types for diagnosing AD. The column for ALL shows the results using all cell type data.

| | Veinous | SMC | Arterial | OPC | Microglia | P. Fibro | Ependymal | Neuron | M. Fibro | T cell | Oligo | Astrocyte | Pericyte | Capillary | ALL |
|---|---|---|---|---|---|---|---|---|---|---|---|---|---|---|---|
| AD 1 | 0.79/817 | 0.91/649 | 0.82/642 | 0.86/98 | 0.72/160 | 0.90/258 | 0.70/50 | 0.68/31 | 0.67/9 | 0.46/35 | 0.83/797 | 0.92/318 | 0.98/1480 | 0.75/1582 | 0.85/6926 |
| AD 2 | 0.61/56 | 0.65/65 | 0.72/64 | 0.89/162 | 0.67/84 | 0.84/62 | 0.96/48 | 0.86/59 | 0.74/38 | 0.80/10 | 0.86/1265 | 0.83/1155 | 0.54/249 | 0.42/144 | 0.79/3461 |
| Control 1 | 0.03/420 | 0.04/361 | 0.08/231 | 0.36/234 | 0.01/95 | 0.27/171 | 0.16/92 | 0.38/58 | 0.31/39 | 0.45/22 | 0.13/634 | 0.07/1730 | 0.01/1471 | 0.02/881 | 0.07/6439 |
| Control 2 | 0.14/604 | 0.05/495 | 0.06/496 | 0.45/130 | 0.08/89 | 0.06/150 | 0.53/99 | 0.22/122 | 0.32/31 | 0.33/27 | 0.36/1184 | 0.43/637 | 0.23/591 | 0.06/504 | 0.23/5159 |
| Single Cell Type Score | 0.808 | 0.866 | 0.849 | 0.733 | 0.826 | 0.851 | 0.743 | 0.735 | 0.693 | 0.617 | 0.800 | 0.810 | 0.821 | 0.773 | 0.835 |

**Supplementary Figure 4: Low-dimensional embeddings of single-cell data used in this paper.** a) UMAP scatter plot of the single-cell data used for our model colored by the 9 AD patients (AD*) and 8 healthy controls (C*). b) UMAP scatter plot colored by 14 cell types.

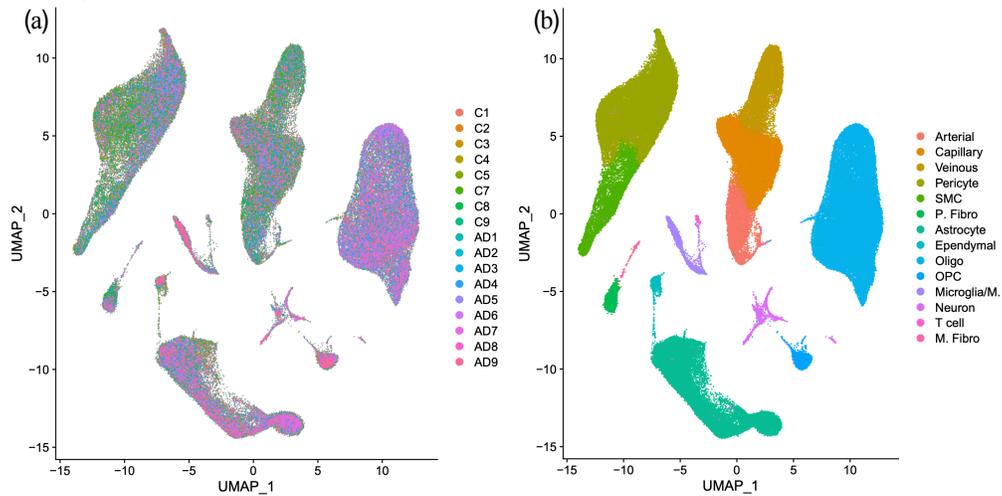

**Supplementary Figure 5. The Attention System diagram.** The attention system with $h$ attention head and $c$ attention subsystems within each attention head.

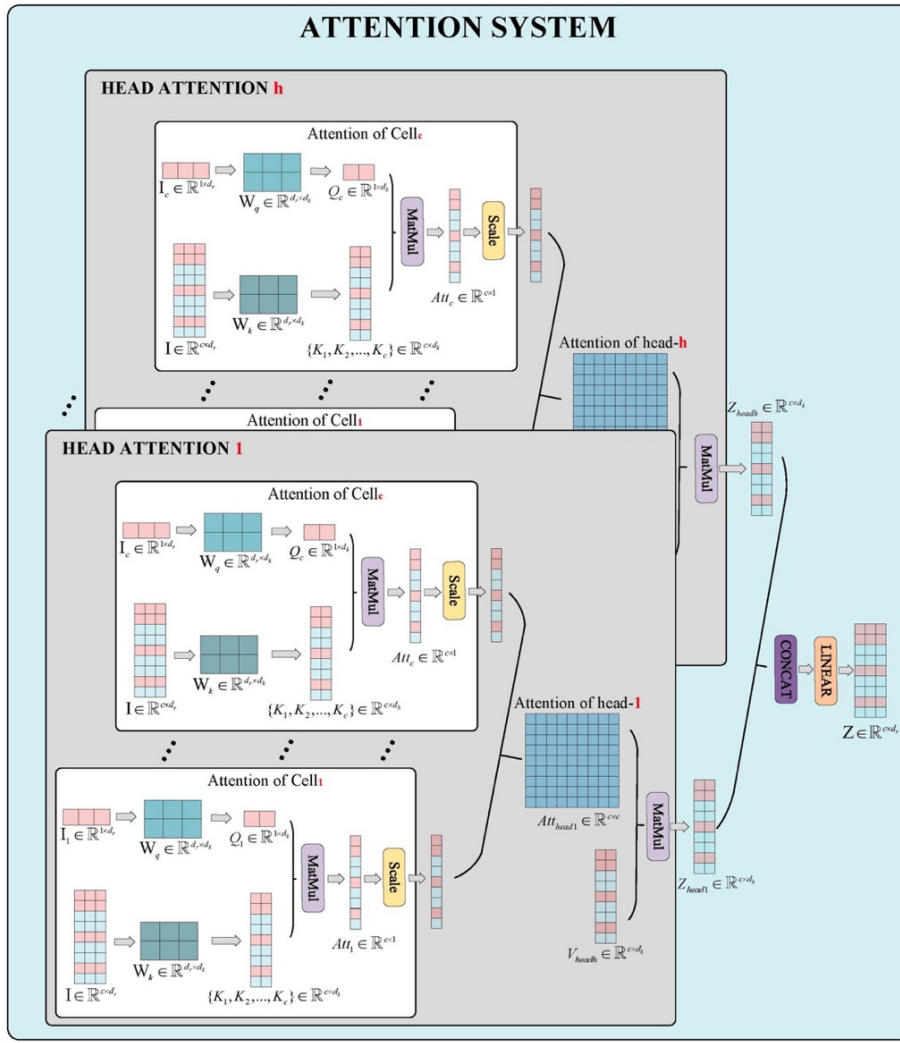

**Supplementary Figure 6. Performances of model with different hyperparameter settings.** The model size and performance of our model with two different hyperparameters are shown as the x-axis and y-axis.

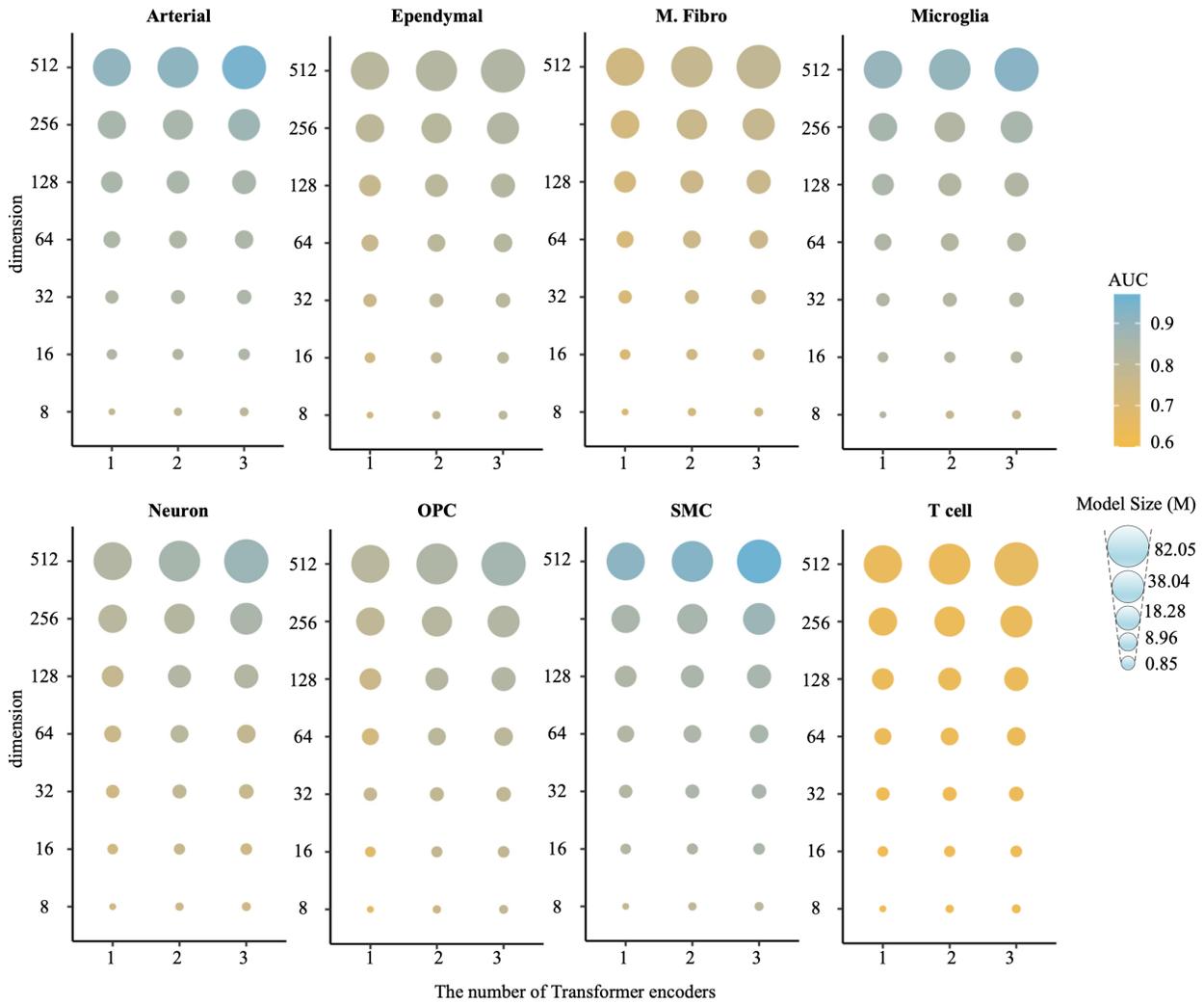

**Supplementary Figure 7. The training process for the baseline models.**

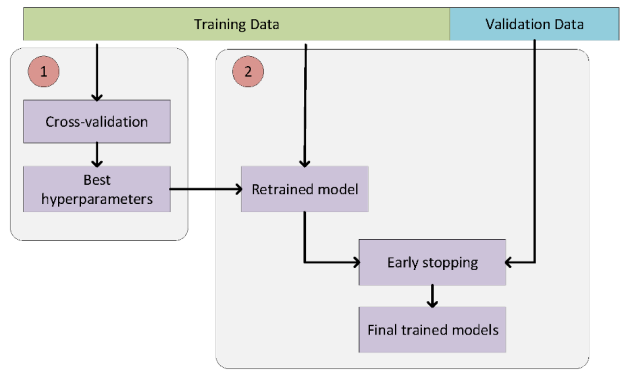